\documentclass{article}

\input{tcilatex}

\begin{document}

\title{Biconformal Matter Actions}
\author{Andr\'{e} Wehner and James T. Wheeler}
\maketitle

\begin{abstract}
We extend $2n$-dim biconformal gauge theory by including Lorentz-scalar
matter fields of arbitrary conformal weight. We show that for a massless
scalar field of conformal weight zero in a torsion-free biconformal
geometry, the solution is determined by the Einstein equation on an $n$-dim
submanifold, with the stress-energy tensor of the scalar field as source.
The matter field satisfies the $n$-dim Klein-Gordon equation.
\end{abstract}

\section{Introduction}

Recently, we developed a new gauge theory of the conformal group, which
solved many of the problems typically associated with scale invariance \cite
{New Conformal Gauging Paper}. In particular, this new class of \textit{%
biconformal geometries} has been shown to resolve the problem of writing
scale-invariant vacuum gravitational actions in arbitrary dimension \textit{%
without} the use of compensating fields \cite{WW}. In the cited work, we
wrote the most general linear vacuum action and completely solved the
resulting field equations subject only to a minimal torsion assumption. We
found that all such solutions were foliated by equivalent $n$-dimensional
Ricci-flat Riemannian spacetimes.

Reference \cite{WW} left an open question: how are matter fields coupled to
biconformal gravity? \textit{A priori}, it is not at all obvious that any
action for biconformal matter permits the same embedded $n$-dim Riemannian
structure that occurs for the vacuum case since biconformal fields are $2n$%
-dimensional. Indeed, in the case of standard $n$-dim conformal gauging (
\cite{Romao+Ferber+Freund}-\cite{Kaku+Townsend2}), we generally require
compensating fields to recover the Einstein equation with matter (see, for
example, \cite{Deser}-\cite{van Nieuwenhuizen}). To answer this question for
biconformal space, in the present work we extend the results of \cite{WW} by
introducing a set of Klein-Gordon-type fields $\phi ^{m}$ of conformal
weight $m$ into the theory. Using the Killing metric intrinsic to
biconformal space, we write the natural kinetic term in the biconformally
covariant derivatives of $\phi ^{m}$ and find the resulting gravitationally
coupled field equations. Then, for the case of one scalar field $\phi $ of
conformal weight zero, we completely solve the field equations, under the
assumption of vanishing torsion. We find that, as before, the solutions are
foliated by equivalent $n$-dim Riemannian spacetime submanifolds whose
curvatures now satisfy the usual Einstein equations with scalar matter. The
field $\phi ,$ which \textit{a priori }depended on all $2n$ biconformal
coordinates, is completely determined by the $n$ coordinates of the
submanifolds and satisfies the submanifold massless Klein-Gordon equation $%
\eta ^{cd}D_{c}D_{d}\phi =0$.

Thus, the new gauging establishes a clear connection between conformal gauge
theory and general relativity with scalar matter, without the use of
compensating fields.

The structure of the paper is as follows. In the next section, we extend the
applicability of the biconformal dual introduced in \cite{WW} and establish
how to write the usual kinetic action for scalar fields of arbitrary
conformal weight using differential forms. Then, in Sec.(3) we find the
field equations resulting from this action coupled to the linear gravity
action introduced in \cite{WW}. Interestingly, these equations together with
previous results show that pairs of scalar matter fields of conjugate
conformal weight provide a source for the Weyl vector. Next, restricting to
the case of a zero-weight scalar field, Secs.(4) and (5) give the solutions
for the curvature and connection, respectively, of the background geometry
in the case of vanishing torsion. Finally, in Sec.(6), we examine the field
equations and other constraints on the matter field.

\section{The biconformal dual and inner product}

For full detail on the new conformal gauging we refer to \cite{New Conformal
Gauging Paper}. We will use the same notation as in \cite{WW}.

The Minkowski metric is written as $\eta _{ab}=$\ $diag(1\ldots 1,-1)$,
where $a,b,\ldots =1,\ldots ,n$. We denote the connection components (gauge
fields) associated with the Lorentz, translation, co-translation, and
dilation generators of the conformal group $O(n,2)$, $n>2$, as the
spin-connection $\mathbf{\omega }_{b}^{a}$, the solder-form $\mathbf{\omega }%
^{a}$, the co-solder-form $\mathbf{\omega }_{a}$, and$\mathbf{\ }$the Weyl
vector $\mathbf{\omega }_{0}^{0}$, respectively. The corresponding $O(n,2)$\
curvatures $\mathbf{\Omega }_{\tilde{b}}^{\tilde{a}}$ ($\tilde{a},\tilde{b}%
,\ldots =0,1,\ldots ,n$) are referred to as the curvature $\mathbf{\Omega }%
_{b}^{a}$,\ torsion $\mathbf{\Omega }^{a}\equiv \mathbf{\Omega }_{0}^{a}$,
co-torsion $\mathbf{\Omega }_{a}\equiv \mathbf{\Omega }_{a}^{0}$, and\
dilation $\mathbf{\Omega }_{0}^{0}$, respectively, and are defined by the
biconformal structure equations, 
\begin{eqnarray}
\mathbf{\Omega }_{b}^{a} &=&\mathbf{d\omega }_{b}^{a}-\mathbf{\omega }%
_{b}^{c}\mathbf{\omega }_{c}^{a}-\Delta _{cb}^{ad}\mathbf{\omega }_{d}%
\mathbf{\omega }^{c}  \label{Structure 1} \\
\mathbf{\Omega }^{a} &=&\mathbf{d\omega }^{a}-\mathbf{\omega }^{b}\mathbf{%
\omega }_{b}^{a}-\mathbf{\omega }_{0}^{0}\mathbf{\omega }^{a}
\label{Structure 2} \\
\mathbf{\Omega }_{a} &=&\mathbf{d\omega }_{a}-\mathbf{\omega }_{a}^{b}%
\mathbf{\omega }_{b}-\mathbf{\omega }_{a}\mathbf{\omega }_{0}^{0}
\label{Structure 3} \\
\mathbf{\Omega }_{0}^{0} &=&\mathbf{d\omega }_{0}^{0}-\mathbf{\omega }^{a}%
\mathbf{\omega }_{a},  \label{Structure 4}
\end{eqnarray}
where $\Delta _{cd}^{ab}\equiv \delta _{c}^{a}\delta _{d}^{b}-\eta ^{ab}\eta
_{cd}$. In all cases differential forms are bold and the wedge product is
assumed between adjacent forms. The position of any lower-case Latin index
corresponds to the associated conformal weight: each upper index contributes 
$+1$ to the weight, while each lower index contributes $-1.$

Biconformal space is the $2n$-dimensional base space of the $O(n,2)$
principal bundle with homothetic fiber, first constructed in \cite{New
Conformal Gauging Paper}. Each biconformal curvature may be expanded in the $%
(\mathbf{\omega }^{a},\mathbf{\omega }_{b})$\ basis as 
\begin{equation}
\mathbf{\Omega }_{\tilde{b}}^{\tilde{a}}=\tfrac{1}{2}\Omega _{\tilde{b}cd}^{%
\tilde{a}}\mathbf{\omega }^{cd}+\Omega _{\tilde{b}d}^{\tilde{a}c}\mathbf{%
\omega }_{c}\mathbf{\omega }^{d}+\tfrac{1}{2}\Omega _{\tilde{b}}^{\tilde{a}%
cd}\mathbf{\omega }_{cd},  \label{Curvterms}
\end{equation}
where we use the convention of writing 
\[
\mathbf{\omega }^{ab\ldots c}\equiv \mathbf{\omega }^{a}\mathbf{\omega }%
^{b}\ldots \mathbf{\omega }^{c}=\mathbf{\omega }^{a}\wedge \mathbf{\omega }%
^{b}\wedge \ldots \wedge \mathbf{\omega }^{c}.
\]
\ The three terms of eq.(\ref{Curvterms}) will be called the spacetime-,
cross-, and momentum-term, respectively, of the corresponding curvature.

Any $r$-form $\mathbf{U}$\ ($r\leq 2n$) defined on the cotangent bundle to
biconformal space can be uniquely decomposed into a sum of $(p,q)$-forms, 
\[
\mathbf{U}=\sum_{p=0}^{r}\mathbf{U}_{p,r-p},
\]
each of which is of the form 
\begin{equation}
\mathbf{U}_{p,q}=\tfrac{1}{p!q!}U_{a_{1}\cdots a_{p}}^{\qquad b_{1}\cdots
b_{q}}\mathbf{\omega }^{a_{1}\cdots a_{p}}\mathbf{\omega }_{b_{1}\cdots
b_{q}}\qquad (p,q\leq n,\;p+q=r)  \label{p-q-form}
\end{equation}
and has conformal weight $p-q$. For example, a $1$-form can be written as 
\begin{equation}
\mathbf{U}=U_{a}\mathbf{\omega }^{a}+U^{a}\mathbf{\omega }_{a}\equiv U_{A}%
\mathbf{\omega }^{A},  \label{1-form}
\end{equation}
where capital Latin indices denote both upper and lower lower-case Latin
indices.

Biconformal space possesses a natural metric, 
\[
K^{AB}=\left( 
\begin{array}{cc}
0 & \delta _{b}^{a} \\ 
\delta _{b}^{a} & 0
\end{array}
\right) 
\]
which is obtained when the non-degenerate Killing form of the conformal
group $O(n,2)$ is restricted to the biconformal base space. The Killing
metric defines a natural inner product between two $1$-forms $\mathbf{U}$
and $\mathbf{V}$: 
\begin{equation}
\left\langle \mathbf{U},\mathbf{V}\right\rangle \equiv \tfrac{1}{2}%
K^{AB}U_{A}V_{B}=\tfrac{1}{2}\left( U^{a}V_{a}+V^{a}U_{a}\right) 
\label{Inner product}
\end{equation}
Notice that, because the metric is essentially $\delta _{b}^{a}$, whenever
we sum an upper with a lower index we have implicitly used the Killing
metric.

In \cite{WW}, we demonstrated that when the indices of the $2n$-dimensional
Levi-Civita symbol are sorted by weight, it may be written as the product of
two $n$-dimensional Levi-Civita symbols of opposite weights: 
\[
\varepsilon _{a_{1}\cdots a_{n}}^{\qquad \quad b_{1}\cdots
b_{n}}=\varepsilon _{a_{1}\cdots a_{n}}^{\qquad \quad }\varepsilon
^{b_{1}\cdots b_{n}}, 
\]
where the mixed index positioning indicates the scaling weight of the
indices, and \underline{not}\ any use of the metric. The Levi-Civita tensor
is normalized such that traces are given by 
\[
\varepsilon _{a_{1}\cdots a_{p}c_{p+1}\cdots c_{n}}\varepsilon ^{b_{1}\cdots
b_{p}c_{p+1}\cdots c_{n}}=p!(n-p)!\delta _{a_{1}\cdots a_{p}}^{b_{1}\cdots
b_{p}}, 
\]
where the antisymmetric $\delta $-symbol is defined as 
\[
\delta _{a_{1}\cdots a_{p}}^{b_{1}\cdots b_{p}}\equiv \delta
_{a_{1}}^{[b_{1}}\cdots \delta _{a_{p}}^{b_{p}]}. 
\]
Using the Levi-Civita tensor, the scale-invariant volume form of biconformal
space is given by: 
\[
\mathbf{\Phi }=\varepsilon _{a_{1}\cdots a_{n}}^{\qquad b_{1}\cdots b_{n}}%
\mathbf{\omega }^{a_{1}\ldots a_{n}}\mathbf{\omega }_{b_{1}\cdots b_{n}}. 
\]
Notice that despite the mixed index positions, $\varepsilon _{a_{1}\cdots
a_{n}}^{\qquad b_{1}\cdots b_{n}}$ is totally antisymmetric on all $2n$
indices.

We now define the biconformal dual of a general $r$-form. Let $\mathbf{U}%
_{p,q}$\ be an arbitrary $(p,q)$-form as denoted in (\ref{p-q-form}). Then
the dual of $\mathbf{U}_{p,q}$\ is an $(n-q,n-p)$-form, also of weight $p-q,$%
\ defined as 
\[
^{\ast }\mathbf{U}_{p,q}\equiv \frac{\tau (p,q)}{p!q!(n-p)!(n-q)!}%
U_{a_{1}\cdots a_{p}}^{\qquad b_{1}\cdots b_{q}}\varepsilon _{b_{1}\cdots
b_{n}}^{\qquad \ a_{1}\cdots a_{n}}\mathbf{\omega }^{b_{q+1}\cdots b_{n}}%
\mathbf{\omega }_{a_{p+1}\cdots a_{n}},
\]
where 
\[
\tau (p,q)=\left\{ 
\begin{array}{c}
1\qquad \qquad \qquad if\qquad p\geq q \\ 
(-1)^{n(p+q)}\qquad if\qquad p<q
\end{array}
\right. 
\]
This choice of $\tau (p,q)$ guarantees that for any $r$-form $\mathbf{U,}$ 
\[
^{\ast \ast }\mathbf{U}=(-1)^{r(n-r)}\mathbf{U}
\]
and that for two arbitrary $(p,q)$- and $(q,p)$-forms, $\mathbf{U}_{p,q}$
and $\mathbf{V}_{q,p}$, respectively, 
\[
\mathbf{U}_{p,q}^{\quad \ast }\mathbf{V}_{q,p}=\mathbf{V}_{q,p}^{\quad \ast }%
\mathbf{U}_{p,q}\mathbf{,}
\]
where we again assume wedge products between forms. It is then easy to show
that for any two $r$-forms $\mathbf{U}$ and $\mathbf{V}$, the product $%
\mathbf{U}^{\ast }\mathbf{V}$ is proportional to the volume form $\mathbf{%
\Phi }$ and 
\[
\mathbf{U}^{\ast }\mathbf{V}=\sum_{p=0}^{r}\mathbf{U}_{p,r-p}^{\qquad \ast }%
\mathbf{V}_{r-p,p}=\sum_{p=0}^{r}\mathbf{V}_{p,r-p}^{\qquad \ast }\mathbf{U}%
_{r-p,p}=\mathbf{V}^{\ast }\mathbf{U,}
\]
because $\mathbf{U}_{p,r-p}^{\qquad \ast }\mathbf{V}_{r-q,q}$ vanishes
unless $p=q$.

Now let $\mathbf{U}$ be a general $1$-form as in (\ref{1-form}). Then 
\[
^{\ast \ast }\mathbf{U}=(-1)^{n-1}\mathbf{U}\ 
\]
and 
\begin{equation}
\tfrac{1}{2}\mathbf{U}^{\ast }\mathbf{U=}\tfrac{(-1)^{n}}{n!^{2}}U_{a}U^{a}%
\mathbf{\Phi .}  \label{U star U}
\end{equation}
Thus, by eq.(\ref{Inner product}), the term $\mathbf{U}^{\ast }\mathbf{U}$
is proportional to the inner product $\left\langle \mathbf{U},\mathbf{U}%
\right\rangle $: 
\begin{equation}
\mathbf{U}^{\ast }\mathbf{U}=\tfrac{2(-1)^{n}}{n!^{2}}\left\langle \mathbf{U}%
,\mathbf{U}\right\rangle \mathbf{\Phi =}\tfrac{(-1)^{n}}{n!^{2}}%
K^{AB}U_{A}U_{B}\mathbf{\Phi .}  \label{U star U 2}
\end{equation}

We are now ready to build the biconformal theory of scalar matter. Let $\phi
^{m}$ be a set of massless Lorentz-scalar fields of conformal weight $m\in 
\mathbf{Z}$ \cite{Extended conformal paper} and $\mathbf{D}\phi ^{m}$ be
their biconformally covariant derivatives defined by 
\begin{equation}
\mathbf{D}\phi ^{m}\equiv \mathbf{d}\phi ^{m}+m\mathbf{\omega }_{0}^{0}\phi
^{m}.  \label{Cov Der}
\end{equation}
This covariant derivative is a $1$-form which is also of weight $m$ and can
be expanded as 
\[
\mathbf{D}\phi ^{m}=\mathbf{\omega }^{A}D_{A}\phi ^{m}.
\]
Since the dual operator preserves the conformal weight, $^{\ast }\mathbf{D}%
\phi ^{-m}$ must be of weight $-m$, so that every term in the infinite sum 
\[
\sum_{m\in \mathbf{Z}}\mathbf{D}\phi ^{m\ast }\mathbf{D}\phi ^{-m}
\]
is of conformal weight zero. Using (\ref{U star U 2}) we see that 
\[
\mathbf{D}\phi ^{m\ast }\mathbf{D}\phi ^{-m}=\tfrac{(-1)^{n}}{n!^{2}}%
K^{AB}D_{A}\phi ^{m}D_{B}\phi ^{-m}\mathbf{\Phi }
\]
for every $m$. We have thus arrived at the Weyl-scalar-valued action 
\begin{equation}
S_{M}=\tfrac{1}{2}\bar{\lambda}\sum_{m}\int \mathbf{D}\phi ^{m\ast }\mathbf{D%
}\phi ^{-m}=\tfrac{1}{2}\bar{\lambda}\tfrac{(-1)^{n}}{n!^{2}}\sum_{m}\int
K^{AB}D_{A}\phi ^{m}D_{B}\phi ^{-m}\mathbf{\Phi .}  \label{Matter}
\end{equation}
We shall make use of the `dual' form of $S_{M}$ when we vary the action with
respect to the field and with respect to the connection, whereas the form of 
$S_{M}$ that explicitly displays the dependence on the Killing metric proves
more useful in varying the base forms.

\section{The linear scalar action}

In a $2n$-dimensional biconformal space the most general Lorentz and
scale-invariant action which is linear in the biconformal curvatures and
structural invariants is 
\[
S_{G}=\int (\alpha \mathbf{\Omega }_{b_{1}}^{a_{1}}+\beta \delta
_{b_{1}}^{a_{1}}\mathbf{\Omega }_{0}^{0}+\gamma \mathbf{\omega }^{a_{1}}%
\mathbf{\omega }_{b_{1}})\mathbf{\omega }^{a_{2}...a_{n}}\mathbf{\omega }%
_{b_{2}...b_{n}}\varepsilon ^{b_{1}...b_{n}}\varepsilon _{a_{1}...a_{n}},
\]
first introduced in \cite{WW}. We will always assume non-vanishing $\alpha ,$%
\ $\beta ,$ and $\gamma $. For a set of massless Lorentz scalar fields $\phi
^{m}$ of weight $m$ with kinetic term $S_{M}$ given by (\ref{Matter}), we
have 
\begin{equation}
S=S_{M}+S_{G}.  \label{Action}
\end{equation}

Variation of this action with respect to the scalar fields yields the
equation 
\begin{equation}
0=\mathbf{D}^{\ast }\mathbf{D}\phi ^{m}  \label{F-Eq 0}
\end{equation}
for every $m$, where 
\[
\mathbf{D}^{\ast }\mathbf{D}\phi ^{m}\equiv \mathbf{d}^{\ast }\mathbf{D}\phi
^{m}+m\mathbf{\omega }_{0}^{0\ast }\mathbf{D}\phi ^{m}.
\]
\bigskip Variation with respect to the connection one-forms gives rise to
the following field equations: 
\begin{eqnarray}
\beta (\Omega ^{a}{}_{ba}-2\Omega _{ca}^{\ d}\delta _{db}^{ca}) &=&-\lambda
\Theta _{b}  \label{F-Eq 1a} \\
\beta (\Omega _{a}^{\ ba}-2\Omega _{\ a}^{cd}\delta _{dc}^{ab}) &=&\lambda
\Theta ^{b}  \label{F-Eq 1b} \\
\alpha (-\Delta _{eg}^{af}\Omega _{\ ab}^{b}+2\Delta _{eb}^{cf}\delta
_{dg}^{ab}\Omega _{ac}^{\ d}) &=&0  \label{F-Eq 2a} \\
\alpha (-\Delta _{eb}^{gf}\Omega _{a}^{\ ab}+2\Delta _{ed}^{af}\delta
_{ab}^{gc}\Omega _{\ c}^{bd}) &=&0  \label{F-Eq 2b} \\
\alpha \Omega _{bac}^{a}+\beta \Omega _{0bc}^{0} &=&-\lambda \Upsilon _{bc}
\label{F-Eq 3a} \\
2(\alpha \Omega _{cd}^{ec}+\beta \Omega _{0d}^{0e})\delta _{eb}^{ad}+\Lambda
_{b}^{a} &=&\lambda \Upsilon _{b}^{a}  \label{F-Eq 3b} \\
\alpha \Omega _{a}^{bac}+\beta \Omega _{0}^{0bc} &=&\lambda \Upsilon ^{bc}
\label{F-Eq 4a} \\
2(\alpha \Omega _{dc}^{ce}+\beta \Omega _{0d}^{0e})\delta _{eb}^{ad}+\Lambda
_{b}^{a} &=&\lambda \Upsilon _{b}^{a}  \label{F-Eq 4b}
\end{eqnarray}
where the matter sources are given by 
\begin{eqnarray*}
\Upsilon _{ab} &\equiv &\sum_{m}D_{a}\phi ^{m}D_{b}\phi ^{-m}=\Upsilon _{ba}
\\
\Upsilon _{b}^{a} &\equiv &\sum_{m}\left( D^{a}\phi ^{m}D_{b}\phi
^{-m}+D^{c}\phi ^{m}D_{c}\phi ^{-m}\delta _{b}^{a}\right)  \\
\Upsilon ^{ab} &\equiv &\sum_{m}D^{a}\phi ^{m}D^{b}\phi ^{-m}=\Upsilon ^{ba}
\\
\Theta _{b} &=&\sum_{m}m\phi ^{m}D_{b}\phi ^{-m} \\
\Theta ^{b} &=&\sum_{m}m\phi ^{m}D^{b}\phi ^{-m}
\end{eqnarray*}
and we have defined 
\begin{eqnarray*}
\Lambda _{b}^{a} &\equiv &(\alpha (n-1)-\beta +\gamma n^{2})\delta _{b}^{a}
\\
\lambda  &\equiv &\frac{1}{(n-1)!^{2}}\bar{\lambda}.
\end{eqnarray*}
Note that since the spin connection does not occur in the covariant
derivative (\ref{Cov Der}), $\delta _{\mathbf{\omega }_{b}^{a}}S_{M}\equiv 0$%
, and there is no matter contribution to eq.(\ref{F-Eq 2a}) or (\ref{F-Eq 2b}%
). Combining equations (\ref{F-Eq 3b}) and (\ref{F-Eq 4b}) we see that the
latter can be replaced by 
\begin{equation}
\Omega _{cd}^{ac}=\Omega _{dc}^{ca}.  \label{F-Eq 4c}
\end{equation}

We remark that the biconformal structure equations together with eqs.(\ref
{F-Eq 1a})-(\ref{F-Eq 2b}) may be used to express the torsion and co-torsion
in terms of the connection, the Weyl vector, and (here) the matter fields $%
\phi ^{m}$. In \cite{WW} it was observed that constraining the torsion to
vanish also forces the Weyl vector to vanish, but this conclusion no longer
holds with matter present. This suggests that setting the torsion to zero is
not an undue constraint as assumed in \cite{WW}, but rather, that the Weyl
vector vanishes unless there are appropriate matter fields present. Taking
this view, we are free to assume $\mathbf{\Omega }^{a}=0.$ Then, there
exists a gauge in which the Weyl vector is given in terms of covariant
derivatives of the fields $\phi ^{m}$ by 
\[
\mathbf{\omega }_{0}^{0}=\frac{1}{(n-1)(n-2)}\frac{\lambda }{\beta }%
\sum_{m}m\phi ^{m}\mathbf{D}\phi ^{-m}, 
\]
Notice that $\mathbf{\omega }_{0}^{0}=0 $ unless conjugate weights, $+m$ and 
$-m$ are both present.

We will explore such dilational sources further elsewhere. Here, since our
goal is to derive the usual form of the Einstein equations with scalar
matter, it is sufficient to restrict our attention to the case $m=0.$ Thus,
for the remainder of this paper we restrict to the case of a scalar field $%
\phi $ of conformal weight zero. Then, the covariant derivative is simply
the exterior derivative 
\[
\mathbf{D}\phi \equiv \mathbf{d}\phi =\mathbf{\omega }^{A}d_{A}\phi ,
\]
so the field equations reduce to 
\begin{eqnarray}
0 &=&^{\ast }\mathbf{d}^{\ast }\mathbf{d}\phi   \label{F-Eq 0a} \\
0 &=&\Omega ^{a}{}_{ba}-2\Omega _{ca}^{\ d}\delta _{db}^{ca}
\label{F-Eq 1a-A} \\
0 &=&\Omega _{a}^{\ ba}-2\Omega _{\ a}^{cd}\delta _{dc}^{ab}
\label{F-Eq 1b-A}
\end{eqnarray}
together with eqs.(\ref{F-Eq 2a})-(\ref{F-Eq 4c}), where $\Theta _{b}=0$, $%
\Theta ^{b}=0$ and 
\begin{eqnarray*}
\Upsilon _{ab} &\equiv &d_{a}\phi \ d_{b}\phi =\Upsilon _{ba} \\
\Upsilon _{b}^{a} &\equiv &d^{a}\phi \ d_{b}\phi +d^{c}\phi \ d_{c}\phi \
\delta _{b}^{a} \\
\Upsilon ^{ab} &\equiv &\ d^{a}\phi \ d^{b}\phi =\Upsilon ^{ba}.
\end{eqnarray*}

\section{Solution for the curvatures}

We now find the most general solution to these equations subject only to the
constraint of vanishing torsion. As discussed above, this condition no
longer implies a vanishing Weyl vector. The Weyl vector nonetheless vanishes
because of our choice to consider only zero conformal weight matter.

Despite vanishing torsion, the general approach to solving the field
equations follows that of \cite{WW}. Starting with a general ansatz for the
spin connection and Weyl vector, we first solve the torsion and co-torsion
equations, eqs.(\ref{F-Eq 1a-A}, \ref{F-Eq 1b-A}) and eqs.(\ref{F-Eq 2a}, 
\ref{F-Eq 2b}). The Bianchi identity following from the vanishing torsion
constraint and field equations (\ref{F-Eq 3a})-(\ref{F-Eq 4c}) then
determines the form of the curvature and dilation. In Sec.(5), we show that
the vanishing torsion constraint also leads to a foliation by $n$%
-dimensional flat Riemannian manifolds. By invoking the gauge freedom on
each of these manifolds, we show the existence of a second foliation by $n$%
-dimensional Riemannian spacetimes satisfying the Einstein equations with
scalar matter.

To begin, we write the spin connection $\mathbf{\omega }_{b}^{a}$ as 
\begin{equation}
\mathbf{\omega }_{b}^{a}=\mathbf{\alpha }_{b}^{a}+\mathbf{\beta }_{b}^{a}+%
\mathbf{\gamma }_{b}^{a}  \label{Spin connection}
\end{equation}
with $\mathbf{\alpha }_{b}^{a}$ and $\mathbf{\beta }_{b}^{a}$ defined by 
\begin{eqnarray*}
\mathbf{d\omega }^{a} &=&\mathbf{\omega }^{b}\mathbf{\alpha }_{b}^{a}+\tfrac{%
1}{2}\Omega ^{abc}\mathbf{\omega }_{bc} \\
\mathbf{d\omega }_{a} &=&\mathbf{\beta }_{a}^{b}\mathbf{\omega }_{b}+\tfrac{1%
}{2}\Omega _{abc}\mathbf{\omega }^{bc}.
\end{eqnarray*}
Using this ansatz in structure equations (\ref{Structure 2}) and (\ref
{Structure 3}), $\Omega ^{abc}$ and $\Omega _{abc}$ remain related to
derivatives of the solder- and co-solder forms, whereas the other torsion
and co-torsion terms are algebraic in the components of $\mathbf{\alpha }%
_{b}^{a},\mathbf{\beta }_{b}^{a}$, $\mathbf{\gamma }_{b}^{a}$, and $\mathbf{%
\omega }_{0}^{0}$. Thus, the separation of the connection allows us to solve
the torsion/co-torsion field equations (\ref{F-Eq 1a-A}), (\ref{F-Eq 1b-A}),
(\ref{F-Eq 2a}), and (\ref{F-Eq 2b}) algebraically.

We simply state the result of this reduction here. More detail is available
in \cite{WW}. Defining 
\[
\mathbf{\sigma }_{b}^{a}\equiv \mathbf{\alpha }_{b}^{a}-\mathbf{\beta }%
_{b}^{a}\equiv \sigma _{bc}^{a}\mathbf{\omega }^{c}+\sigma _{b}^{ac}\mathbf{%
\omega }_{c}
\]
and setting 
\[
\mathbf{\Omega }^{a}=0,
\]
field equations (\ref{F-Eq 1a-A}), (\ref{F-Eq 1b-A}), (\ref{F-Eq 2a}), and (%
\ref{F-Eq 2b}) imply 
\begin{eqnarray*}
\mathbf{\omega }_{0}^{0} &=&0 \\
\sigma _{bc}^{a} &=&0 \\
\sigma _{a}^{ba} &=&0
\end{eqnarray*}
with no assumption concerning the co-torsion, curvature, or dilation. From
these we find 
\[
\mathbf{\omega }_{b}^{a}=\mathbf{\alpha }_{b}^{a}.
\]
The co-torsion cross- and momentum terms reduce to 
\begin{eqnarray*}
\Omega _{ac\ }^{\ b} &=&0 \\
\Omega _{a}^{\ bc} &=&\sigma _{a}^{bc}-\sigma _{a}^{cb},
\end{eqnarray*}
so that the full co-torsion is

\[
\mathbf{\Omega }_{a}=\tfrac{1}{2}\Omega _{acd}\mathbf{\omega }^{cd}+\sigma
_{a}^{bc}\mathbf{\omega }_{bc}
\]
with 
\[
\sigma _{a}^{ba}=0.
\]
The spacetime co-torsion $\Omega _{abc}$ remains undetermined.

Next, we turn our attention to the curvature and dilation equations, eqs.(%
\ref{F-Eq 3a})-(\ref{F-Eq 4c}). The vanishing torsion constraint makes it
possible to obtain an algebraic condition on the curvatures from the Bianchi
identity associated with eq.(\ref{Structure 2}). Taking the exterior
derivative of eq.(\ref{Structure 2}) gives 
\begin{equation}
\mathbf{\omega }^{b}\mathbf{\Omega }_{b}^{a}=\mathbf{\omega }^{a}\mathbf{%
\Omega }_{0}^{0},  \label{Bianchi}
\end{equation}
which implies for the curvature components 
\begin{eqnarray}
\Omega _{0}^{0cd} &=&0  \label{M Dilation} \\
\Omega _{b}^{acd} &=&0  \label{M Curvature} \\
\Omega _{0d}^{0c} &=&-\tfrac{1}{n-1}\Omega _{da}^{ac}
\label{CT Dilation/Curvature} \\
\Omega _{cd}^{ab} &=&\tfrac{1}{n-1}\Delta _{cd}^{fa}\Omega
_{fe}^{eb}=-\Delta _{cd}^{fa}\Omega _{0f}^{0b}  \label{CT Curvature} \\
\Omega _{0cd}^{0} &=&\tfrac{1}{n-2}(\Omega _{dac}^{a}-\Omega _{cad}^{a})
\label{ST Dilation/Curvature}
\end{eqnarray}
Next, we impose field equations (\ref{F-Eq 3a})-(\ref{F-Eq 4c}) onto these
conditions and see that eq. (\ref{F-Eq 4a}) is satisfied by virtue of eqs.(%
\ref{M Dilation}) and (\ref{M Curvature}) if and only if 
\[
\Upsilon ^{ab}=0,
\]
so that 
\begin{eqnarray*}
d^{a}\phi  &=&0 \\
\Upsilon _{b}^{a} &=&0.
\end{eqnarray*}
Imposing eqs.(\ref{F-Eq 3b}) and (\ref{F-Eq 4b}) onto (\ref{CT
Dilation/Curvature}) now yields for $\beta \neq (n-1)\alpha $ 
\[
\Omega _{0b}^{0a}=\tfrac{1}{\alpha (n-1)-\beta }(\lambda (-d^{a}\phi
d_{b}\phi +\tfrac{2}{n-1}d^{c}\phi d_{c}\phi \delta _{b}^{a})-\tfrac{1}{n-1}%
(\alpha (n-1)-\beta +\gamma n^{2})\delta _{b}^{a}
\]
Since $d^{a}\phi =0,$ this reduces to 
\begin{equation}
\Omega _{0b}^{0a}=-\;\frac{(\alpha (n-1)-\beta +\gamma n^{2})}{(n-1)(\alpha
(n-1)-\beta )}\delta _{b}^{a}\equiv -\chi \delta _{b}^{a}.
\label{CT Dilation}
\end{equation}
Imposing eq.(\ref{F-Eq 3a}) onto (\ref{ST Dilation/Curvature}) yields 
\[
\left. 
\begin{array}{c}
\Omega _{0bc}^{0}=0 \\ 
\Omega _{bac}^{a}=\Omega _{cab}^{a} \\ 
\alpha \Omega _{bac}^{a}=-\lambda \Upsilon _{bc}
\end{array}
\right. \qquad if\qquad \beta \neq \tfrac{1}{2}(n-2)\alpha 
\]
If the first of these conditions is substituted back into (\ref{Bianchi}),
we obtain the cyclic identity on the spacetime curvature: 
\[
\Omega _{\lbrack bcd]}^{a}=0.
\]

Summarizing the forms of the dilation and curvature so far we have for the
generic case

\begin{eqnarray}
\mathbf{\Omega }_{0}^{0} &=&-\chi \mathbf{\omega }_{a}\mathbf{\omega }^{a}
\label{Dilation} \\
\mathbf{\Omega }_{b}^{a} &=&\tfrac{1}{2}\Omega _{bcd}^{a}\mathbf{\omega }%
^{cd}+\chi \Delta _{bd}^{ca}\mathbf{\omega }_{c}\mathbf{\omega }^{d},
\label{Curvature}
\end{eqnarray}
with 
\begin{eqnarray}
\Omega _{\lbrack bcd]}^{a} &=&0 \\
\alpha \Omega _{bac}^{a} &=&-\lambda \Upsilon _{bc}.  \label{Ricci}
\end{eqnarray}
We now use the vanishing of the Weyl vector to obtain further constraints on 
$\mathbf{\Omega }_{0}^{0}$ and $\mathbf{\Omega }_{b}^{a}$. Substituting the
restricted form of the dilation (\ref{Dilation}) into structure equation (%
\ref{Structure 4}), 
\[
\mathbf{\Omega }_{0}^{0}=\Omega _{0b}^{0a}\mathbf{\omega }_{a}\mathbf{\omega 
}^{b}=\mathbf{\omega }_{a}\mathbf{\omega }^{a},
\]
we see that 
\begin{equation}
\Omega _{0b}^{0a}=\delta _{b}^{a}.  \label{CT Dilation 2}
\end{equation}
The last equation has to be equal to the dilation crossterm as given by eq.(%
\ref{CT Dilation}), which implies $\chi =-1$, i.e. a relationship between
the constants $\alpha $, $\beta $, and $\gamma $: 
\[
\frac{\gamma n}{\alpha (n-1)-\beta }=-1.
\]
Hence, the volume ($\gamma $) term must necessarily be present in the
action. Thus, in the generic case, where 
\[
\beta \neq (n-1)\alpha ;\;\beta \neq \tfrac{1}{2}(n-2)\alpha ,
\]
we have a two-parameter class of allowed actions, differing only in the
constant $\frac{\lambda }{\alpha }$. Through eq.(\ref{CT Curvature}), eq.(%
\ref{CT Dilation 2}) also implies that 
\[
\Omega _{bd}^{ac}=-\Delta _{db}^{ac}.
\]

We have now satisfied all of field equations (\ref{F-Eq 1a})-(\ref{F-Eq 4c}%
). The curvatures take the form\bigskip 
\begin{eqnarray}
\mathbf{\Omega }^{a} &=&0  \label{Torsion A} \\
\mathbf{\Omega }_{a} &=&\sigma _{a}^{bc}\mathbf{\omega }_{bc}+\tfrac{1}{2}%
\Omega _{abc}\mathbf{\omega }^{bc}  \label{Co-Torsion A} \\
\mathbf{\Omega }_{0}^{0} &=&\mathbf{\omega }_{a}\mathbf{\omega }^{a}
\label{Dilation A} \\
\mathbf{\Omega }_{b}^{a} &=&\tfrac{1}{2}\Omega _{bcd}^{a}\mathbf{\omega }%
^{cd}-\Delta _{bd}^{ca}\mathbf{\omega }_{c}\mathbf{\omega }^{d}
\label{Curvature A}
\end{eqnarray}
subject to the constraints 
\begin{eqnarray*}
\Omega _{\lbrack bcd]}^{a} &=&0 \\
\alpha \Omega _{bac}^{a} &=&-\lambda \Upsilon _{bc} \\
\sigma _{a}^{ba} &=&0 \\
d^{a}\phi  &=&0
\end{eqnarray*}
Notice that the dilation is necessarily non-degenerate, but may not be
closed.

In the next section, we find further constraints on the curvatures arising
from the structure equations.

\section{Solution for the connection}

While eqs.(\ref{Torsion A})-(\ref{Curvature A}) for the curvatures satisfy
all of the field equations, they do not fully incorporate the form of the
biconformal structure equations as embodied in the Bianchi identities.
Therefore, in this section, we turn to the consequences of the form of the
curvatures on the connection.

So far, we have established that in the as yet unspecified original gauge
the Weyl vector vanishes and the spin connection is fully determined by the
solder form $\mathbf{\omega }^{a}$: 
\begin{eqnarray*}
\mathbf{\omega }_{0}^{0} &=&0 \\
\mathbf{\omega }_{b}^{a} &=&\mathbf{\alpha }_{b}^{a}=\alpha _{bc}^{a}\mathbf{%
\omega }^{c}+\alpha _{b}^{ac}\mathbf{\omega }_{c}.
\end{eqnarray*}
Substituting the reduced curvatures into eqs.(\ref{Structure 1})-(\ref
{Structure 3}) (eq.(\ref{Structure 4}) is identically satisfied by (\ref
{Dilation A})), the structure equations now take the form 
\begin{eqnarray}
\mathbf{d\alpha }_{b}^{a} &=&\mathbf{\alpha }_{b}^{c}\mathbf{\alpha }%
_{c}^{a}+\tfrac{1}{2}\Omega _{bcd}^{a}\mathbf{\omega }^{cd}
\label{Structure 1a} \\
\mathbf{d\omega }^{a} &=&\mathbf{\omega }^{b}\mathbf{\alpha }_{b}^{a}
\label{Structure 2a} \\
\mathbf{d\omega }_{a} &=&\mathbf{\alpha }_{a}^{b}\mathbf{\omega }_{b}+\sigma
_{a}^{bc}\mathbf{\omega }_{bc}+\tfrac{1}{2}\Omega _{abc}\mathbf{\omega }%
^{bc}.  \label{Structure 3a}
\end{eqnarray}

We observe that eq.(\ref{Structure 2a}) is in involution. By the Frobenius
theorem, we can consistently set $\mathbf{\omega }^{a}$ to zero and obtain a
foliation by submanifolds, where the spin connection and the co-solder form
reduce to 
\begin{eqnarray*}
\mathbf{f}_{a} &\equiv &\mathbf{\omega }_{a}|_{\mathbf{\omega }^{a}=0} \\
\mathbf{\hat{\alpha}}_{b}^{a} &\equiv &\mathbf{\alpha }^{a}_{b}|_{\mathbf{%
\omega }^{a} =0}=\alpha_{b}^{ac}\mathbf{f}_{c} \\
\hat{\sigma}_{a}^{bc} &\equiv &\sigma^{bc}_{a}|_{\mathbf{\omega}^{a}=0}.
\end{eqnarray*}
Then each submanifold is described by the reduced structure equations 
\begin{eqnarray*}
\mathbf{d\hat{\alpha}}_{b}^{a} &=&\mathbf{\hat{\alpha}}_{b}^{c}\mathbf{\hat{%
\alpha}}_{c}^{a} \\
\mathbf{df}_{a} &=&\mathbf{\hat{\alpha}}_{a}^{b}\mathbf{f}_{b}+\hat{\sigma}%
_{a}^{bc}\mathbf{f}_{bc}.
\end{eqnarray*}
Since the spin-connection is involute, there exists a Lorentz gauge
transformation such that $\mathbf{\hat{\alpha}}_{b}^{a}=0$ on each
submanifold, i.e. $\alpha _{b}^{ac}=0$. With this gauge choice the system
reduces to simply 
\begin{equation}
\mathbf{df}_{a}=\hat{\sigma}_{a}^{bc}\mathbf{f}_{bc}.  \label{df}
\end{equation}
This can be solved in the usual way giving $\hat{\sigma}_{a}^{bc}$ in terms
of $\mathbf{f}_{a}$ and $\mathbf{df}_{a}$. Since this solution has the same
form on each leaf of the foliation, the expression for $\sigma_{a}^{bc}$
remains valid when it is extended back to the full space, i.e. $%
\sigma_{a}^{bc}$ depends on the $2n$ biconformal coordinates only through
its dependence on $\mathbf{f}_{a}$.

The existence of an $\mathbf{\hat{\alpha}}_{b}^{a}=0$ gauge depends only on
vanishing torsion, which leads to an involution of the solder form $\mathbf{%
\omega}^{a}$ and the resulting Bianchi identity, eq.(\ref{Bianchi}).
Therefore, the results of Sec.(4) remain valid in the $\mathbf{\hat{\alpha}}%
_{b}^{a}=0$ gauge.

Returning to the full biconformal space, we now have a gauge 
\[
\mathbf{\omega }^{a}\equiv \mathbf{e}^{a}
\]
such that the spin connection is 
\[
\mathbf{\alpha }_{b}^{a}=\alpha _{bc}^{a}\mathbf{e}^{c},
\]
while the co-solder form may be written in terms of $\mathbf{f}_{a}$ and an
additional term linear in the solder form, 
\begin{equation}
\mathbf{\omega }_{a}=\mathbf{f}_{a}+h_{ab}\mathbf{e}^{b},  \label{co-solder}
\end{equation}
Notice that while $\mathbf{f}_{a}$ depends on all $2n$ coordinates of this
extension, it remains independent of the $1$-forms $\mathbf{e}^{a}$. This
means that $\mathbf{df}_{a}$ remains at least linear in $\mathbf{f}_{a},$
and is consequently involute. We can therefore turn the problem around,
setting $\mathbf{f}_{a}=0$ to obtain a second foliation of the biconformal
space. We can define $\mathbf{h}_{a}$ in terms of this involution, setting 
\[
\mathbf{h}_{a}\equiv \mathbf{\omega }_{a}|_{\mathbf{f}_{a}=0}=h_{ab}\mathbf{e%
}^{b},
\]
with $h_{ab}$ arbitrary. The structure equations for the $\mathbf{f}_{a}=0$
geometry are 
\begin{eqnarray}
\mathbf{d\alpha }_{b}^{a} &=&\mathbf{\alpha }_{b}^{c}\mathbf{\alpha }%
_{c}^{a}+\tfrac{1}{2}\Omega _{bcd}^{a}\mathbf{e}^{cd}  \label{red-alpha} \\
\mathbf{de}^{a} &=&\mathbf{e}^{b}\mathbf{\alpha }_{b}^{a}  \label{red-solder}
\\
\mathbf{dh}_{a} &=&\alpha _{ac}^{b}\mathbf{e}^{c}\mathbf{h}_{b}+\sigma
_{a}^{bc}\mathbf{h}_{bc}+\tfrac{1}{2}\Omega _{abc}\mathbf{e}^{bc}
\label{red-h}
\end{eqnarray}

Eq.(\ref{red-h}) determines $\mathbf{h}_{a}$ once the spacetime co-torsion, $%
\Omega _{abc},$ is given, with little consequence for the rest of the
geometry. We focus our attention on the first two equations. Since they are
unchanged from their full biconformal form, the curvature 
\[
\mathbf{R}_{b}^{a}\equiv \tfrac{1}{2}\Omega _{bcd}^{a}\mathbf{e}^{cd}
\]
and connection $\mathbf{\alpha }_{b}^{a}$ (and of course $\mathbf{e}^{a}$,
by the first involution) are fully determined on the $n$-dimensional $%
\mathbf{f}_{a}=0$ submanifold. Thus, $\mathbf{\alpha }_{b}^{a}$ is the usual
spin connection compatible with $\mathbf{e}^{a}$, while $\mathbf{R}_{b}^{a}$
is its curvature. If we let 
\[
R_{ab}\equiv \Omega _{acb}^{c},
\]
then the Bianchi identity following from (\ref{red-alpha}), 
\[
\mathbf{DR}_{b}^{a}=0,
\]
implies that the tensor 
\[
G_{ab}\equiv R_{ab}-\tfrac{1}{2}\eta _{ab}R
\]
is divergence-free.

Now that $\Omega _{acb}^{c}$ is seen to be the Ricci tensor of an underlying 
$n$-dim submanifold, it follows from the remaining condition (\ref{Ricci}), 
\begin{eqnarray*}
R_{ab} &=&\kappa \Upsilon _{ab} \\
\kappa  &\equiv &-\tfrac{\lambda }{\alpha }
\end{eqnarray*}
that these submanifolds satisfy the Einstein equations, 
\begin{equation}
G_{ab}=\kappa T_{ab},  \label{Einstein}
\end{equation}
with the divergence-free stress-energy tensor given by derivatives of the
matter field: 
\begin{equation}
T_{ab}\equiv d_{a}\phi d_{b}\phi -\tfrac{1}{2}\eta _{ab}\eta ^{ce}d_{c}\phi
d_{e}\phi .  \label{Stress-Energy}
\end{equation}
Even though the co-torsion has a nonvanishing spacetime projection, the
curvature is the one computed from the solder form $\mathbf{\omega }^{a}$
alone. This is our principal result, establishing a direct connection
between general relativity with scalar matter and the more general structure
of biconformal gauge theory with scalar matter.

\section{Constraints on the matter field}

We have seen that the Bianchi identity associated with eq.(\ref{Structure 2}%
) with vanishing torsion together with field equation (\ref{F-Eq 4a}) imply 
\[
d^{a}\phi =0 
\]
or 
\begin{equation}
\mathbf{d}\phi =\mathbf{\omega }^{a}d_{a}\phi .  \label{d-phi}
\end{equation}
We now show that this implies that the scalar field $\phi $ depends only on
the $n$ coordinates spanning each leaf of the $\mathbf{f}_{a}=0$ foliation
and identically satisfies its own field equation.

Based on the involution for $\mathbf{\omega }^{a}$ there exist $n$
coordinates $x^{\mu }$ of weight $+1$ such that 
\[
\mathbf{\omega }^{a}=e_{\mu }^{\;a}\mathbf{d}x^{\mu } 
\]
with the component matrices $e_{\mu }^{\;a}$ necessarily invertible. From
eq.(\ref{Structure 2a}), we immediately find that $e_{\mu }^{\;a}=e_{\mu
}^{\;a}(x).$ Similarly, it can be shown from the $\mathbf{f}_{a}$ involution
that there exist $n$ complementary coordinates $y_{\nu }$ of weight $-1$
such that $\mathbf{f}_{a}$ takes the form 
\[
\mathbf{f}_{a}=f_{a}^{\;\mu }\mathbf{d}y_{\mu }. 
\]
Since both $\mathbf{f}_{a}$ and $\mathbf{d}y_{\mu }$ completely span the
co-tangent bundles of the $\mathbf{\omega }^{a}=0$ submanifolds, the
component matrices $f_{a}^{\;\mu }$ are also necessarily invertible. Thus, $%
(x^{\mu },y_{\mu })$ form a complete set of local coordinates on biconformal
space. If we expand $\mathbf{d}\phi $ in this coordinate basis, we obtain 
\[
\mathbf{d}\phi =\partial _{\mu }\phi \mathbf{d}x^{\mu }+\partial ^{\mu }\phi 
\mathbf{d}y_{\mu }, 
\]
where $(\partial _{\mu },\partial ^{\mu })$ denote derivatives with respect
to $(x^{\mu },y_{\nu })$. Equating this general form with the derived form (%
\ref{d-phi}), 
\[
\mathbf{d}\phi =e_{\mu }^{\;a}(x)d_{a}\phi \mathbf{d}x^{\mu }, 
\]
we see that $\phi =\phi (x)$. Hence, the scalar field is entirely defined on
the $n$-dimensional Riemannian spacetime.

Using eq.(\ref{co-solder}) to expand the co-solder-form as 
\[
\mathbf{\omega }_{a}=f_{a}^{\;\mu }\mathbf{d}y_{\mu }+h_{ab}e_{\mu }^{\;b}%
\mathbf{d}x^{\mu }, 
\]
it is now easy to show that 
\[
d^{a}\phi =0\Rightarrow d^{a}d_{a}\phi =0. 
\]
This result, together with the vanishing torsion constraint and eqs.(\ref
{Structure 2a})-(\ref{Structure 3a}), imply that the field equation for $%
\phi $, eq.(\ref{F-Eq 0a}), is identically satisfied.

However, the field is constrained by the fact that the stress-energy tensor (%
\ref{Stress-Energy}) is, by eq.(\ref{Einstein}), proportional to the
divergence-free Einstein tensor and hence must be itself divergence-free.
Since 
\begin{eqnarray*}
\eta ^{be}D_{e}T_{ab} &=&\eta ^{be}D_{e}\left( D_{a}\phi D_{b}\phi -\tfrac{1%
}{2}\eta _{ab}\eta ^{cf}D_{c}\phi D_{f}\phi \right)  \\
&=&\eta ^{cd}(D_{c}D_{d}\phi )D_{a}\phi 
\end{eqnarray*}
the stress-energy tensor is divergence-free if and only if 
\[
\eta ^{cd}D_{c}D_{d}\phi =0.
\]
This establishes that the scalar field $\phi $ satisfies the free
Klein-Gordon (wave) equation on the $n$-dimensional spacetime.

\section{Conclusions}

We have developed aspects of the theory of scalar matter in biconformal
space. Using the existence of an inner product of $1$-forms and a dual
operator, we constructed an action for a scalar matter field $\phi ^{m}$
coupled to gravity and found the field equations. We solved them for the
case of a scalar field of conformal weight zero in a torsion-free
biconformal geometry. As in the vacuum case, the generic solutions are
foliated by equivalent $n$-dimensional Riemannian spacetime manifolds. The
curvature of each submanifold satisfies the usual Einstein equations with
scalar matter. The scalar field is entirely defined on the submanifold and
satisfies the $n$-dimensional massless Klein-Gordon equation.

\end{document}